     [1999/12/01 v1.4c Il Nuovo Cimento]
\documentclass{cimento}

\usepackage{graphicx}  
\title{Observing Hawking radiation in Bose-Einstein condensates \\ via correlation measurements}
\author{A.~Fabbri\from{ins:x}
}
\instlist{\inst{ins:x} Museo Storico della Fisica e Centro Studi e Ricerche 'Enrico Fermi', 
Piazza del Viminale 1, 00184 Roma, Italy; 
Dipartimento di Fisica dell'Universit\`a di Bologna,
Via Irnerio 46, 40126 Bologna, Italy;
Departamento de F\'isica Te\'orica and IFIC, Universidad de Valencia-CSIC,  
C. Dr. Moliner 50, 46100 Burjassot, Spain
}
\PACSes{\PACSit{04.62.+v}
\PACSes{04.70.Dy, }
{03.75.Kk}
}

\begin{document}

\maketitle

\begin{abstract}
Observing quantum particle creation by black holes (Hawking radiation) in the astrophysical context is, in ordinary situations, hopeless.
Nevertheless the Hawking effect, which depends only on kinematical properties of wave propagation
in the presence of horizons, is present also in nongravitational contexts, for instance in stationary
fluids undergoing supersonic flow. We present results on how to observe the analog Hawking radiation
in Bose-Einstein condensates by a direct measurement of the density correlations due to the phonon pairs (Hawking quanta-partner) created by the acoustic horizon. 
\end{abstract}

\section{Hawking radiation and experimental prospects}

In 1974 Stephen Hawking~\cite{hawking} discovered that due to
quantum mechanics black holes are not ``black", but emit particles
in the form of thermal radiation at a characteristic temperature
\begin{equation}
\label{hawktemp}
T_H=\frac{\hbar\kappa}{2\pi k_B}\ ,
\end{equation}
where $\kappa$ ($=c^3/4GM$ for a Schwarzschild black hole) is the horizon's surface gravity. This
astonishing result was obtained by studying the vacuum of matter fields in the
vicinity of the event horizon, the black hole's boundary marking 
the region from where nothing (not even light) can escape.

Quantum vacuum fluctuactions are usually pictured in terms of pairs
of virtual particles being continuously created out of the vacuum,
but which almost instantaneously (after a time of the order of
Planck time $t_P\sim 10^{-43}\ s$) annihilate and disappear leaving
no observable effect. The presence of the event horizon can change
dramatically this picture, since when one member of the pair is
created just inside the horizon and the other just outside, the one
in the interior (the partner) gets trapped and may leave the other
free to propagate far away where it is detected as a real particle
(Hawking quanta). This heuristic argument was given by Hawking
in~\cite{hawking2} to explain the basic mechanism
responsible for the black hole's emission.

Because of its physical consequences (namely, the information loss problem \cite{hawking3}) 
the Hawking effect is nowadays considered as a milestone for the construction of a 
quantum theory of gravity. Nevertheless, it is practically impossible to observe it. 
Since we cannot, for obvious reasons, check at the same time the
existence of the Hawking quanta and of their partners inside the
horizon, an experimental verification of the Hawking effect requires
the detection of the emitted thermal flux at the temperature
(\ref{hawktemp}) far from the black hole. The bad news is that, unfortunately, in ordinary
situations where the black holes are created from gravitational
collapse of massive stars the Hawking flux at $T_H\sim 10^{-7} K\frac{M_{Sun}}{M} \leq
10^{-7}\ K$ gets completely overwhelmed by the cosmic microwave
background radiation at $T_{CMB}\sim 3\ K$.\footnote{Alternative
possibilities to detect it are based on mini black holes (with a much higher $T_H$) production, 
by density fluctuatons in the early Universe \cite{carr} and in particle accelerators due to the existence
of large extra dimensions \cite{addrs}. }

\section{A way out: analog Hawking radiation from acoustic black holes in fluids} 

To remedy this disappointing situation, in 1981 Bill Unruh \cite{unruh} explored an alternative route.
It was known that the propagation of sound waves in inhomogeneous eulerian fluids and that of (massless) scalar fields in curved backgrounds are mathematically equivalent. Moreover, sound waves get trapped in regions of supersonic flow in the same way light is trapped inside a gravitational black hole. Therefore a stationary fluid configuration consisting
in a region of subsonic flow  and one of supersonic flow connected through an acoustic horizon 
represents an acoustic analog of a gravitational black hole, see Fig. 1.   

\begin{figure}
\centering \includegraphics[angle=0, height=2.0in] {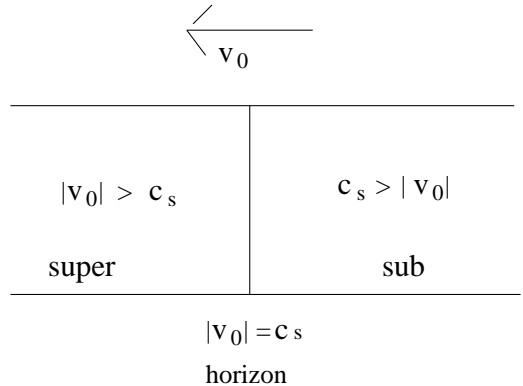}
\caption{Schematic representation of an acoustic black hole in a fluid. $v_0=v_0(x)$ and $c_s=c_s(x)$ are, respectively, the fluid and sound velocities. The left region has supersonic flow and the right one is subsonic; they are connected through an acoustic horizon.}
\end{figure}

Unruh pushed this analogy further at the quantum level. Quantized sound waves (phonons) in these acoustic black holes behave like quantized scalar fields in gravitational black holes. The same analysis Hawking used can then be applied here leading to the prediction that acoustic black holes will produce a thermal flux of phonons from their acoustic horizon at a temperature given by (for a $1D$ fluid 
for simplicity) 
\begin{equation}
\label{hawktempan}T_H^{an}=\frac{1}{4\pi k_Bc_s} \frac{d(c_s^2-v_0^2)}{dx}|_{hor}\ ,
\end{equation}
where $c_s$ is the sound speed, $v_0$ the fluid velocity and the derivative is evaluated at the acoustic horizon 
(where $c_s=|v_0|$).

Although we limited here to the case of fluids, these same considerations apply also to other condensed matter systems in the long wavelength (hydrodynamic) approximation \cite{livrev}.  Besides offering new unexpected possibilities to test the Hawking effect in a nongravitational domain, this application also allows to address from first principles a delicate point in Hawking's derivation called the transplanckian problem. 

It is well known that in the propagation from the horizon to infinity light signals suffer an infinite exponential redshift. Applied to Hawking's analysis, this implies that in order for the Hawking quanta to reach the asymptotic region (where they are detected) the relevant field modes in the near-horizon region must be characterized by an enormous initial frequency, well beyond the Planck scale. This poses a problem because in these regimes one should more appropriately use a quantum theory of gravity, the lack of which poses serious concerns both on the validity of Hawking's analysis and its final result.

One can address this problem in the fluid case, since the theory describing scales below the interatomic distance (the analogous of Planck scale in gravity) is in many cases well known. 

Many directions are being explored, the most recent ones ranging from water tanks experiments (stimulated Hawking radiation from white hole flows \cite{silke}) to laser pulse filaments \cite{faccio}  \footnote{The interpetation of these results is still controversial, see for instance \cite{unshu}.} and to Bose-Einstein condensates (BECs), the focus of this contribution. 

\section{The  gravitational analogy in BECs}

Bose-Einstein condensates are ultracold bosonic systems characterized by the fact that all constituents share the same quantum state. Perfect condensates exist only at absolute zero temperature, while in practice a small non condensed fraction is always present.  To describe them one writes down the Bose field operator $\hat\Psi= \Psi_0(1 + \hat\phi)$, where the classical wavefunction $\Psi_0$ describes the macroscopic condensate and the quantum operator $\hat\psi$ the small noncondensed fraction.  In the dilute gas approximation (see for instance \cite{pistri}) $\Psi_0$ satisfies the Gross-Pitaevski equation 
\begin{equation}\label{gp}
i\hbar\partial_t \Psi_0 = \left(-\frac{\hbar^2\vec \nabla^2}{2m} + V_{ext} + g |\Psi_0|^2 \right)\Psi_0\ ,
\end{equation}
where $m$ is the constituents mass, $V_{ext}$ the trapping potential and $g$ the nonlinear atom-atom interaction constant, 
while $\hat\phi$ treated as a linear perturbation follows the Bogoliubov-de Gennes equation
\begin{equation}\label{bdg}
i\hbar  \partial_t   \hat \phi= - \left( \frac{\hbar^2\vec \nabla^2}{2m} + \frac{\hbar^2}{m}\frac{\vec\nabla \Psi_0 }{\Psi_0} \vec\nabla \right)\hat\phi +mc_s^2 (\hat\phi + \hat\phi^{\dagger})\ ,
\end{equation}
with $c_s=\sqrt{\frac{gn}{m}}$ the speed of sound. Eq. (\ref{gp}) gives the background while (\ref{bdg}) describes the quantum fluctuations and is valid at all scales. 

To make contact with the gravitational analogy we need to consider the hydrodynamical approximation of the above equations, and this is more easily achieved in the density-phase representation in which we write
$\hat \Psi=\sqrt{\hat n}e^{i\hat \theta}$. Again we consider the splitting $\hat n=n_0 +\hat n_1$ and $\hat\theta=\theta_0 +\hat\theta_1$ with $n_0, \theta_0$ the background and $\hat n_1, \hat\theta_1$ the linear fluctuations. Provided the background varies on scales bigger than the healing length $\xi=\frac{\hbar}{mc_s}$ (the microscopic scale, analogous to the Planck length in gravity), eq. (\ref{gp}) for 
$n_0,\ \theta_0$  reduces to the continuity and Euler equations for irrotational inviscid fluids (the analogous of Einstein's field equations). Similar arguments applied to  (\ref{bdg}) give  two  operatorial equations for the density and phase fluctuations 
$\hat n_1=n_0(\hat \phi + \hat \phi^\dagger),\ \hat\theta_1=\frac{\hat \phi - \hat \phi^\dagger}{2i}$ in the form 
\begin{equation}\label{df}
\hat n_1= -\frac{n}{mc_s^2}\left[ v_0\partial_x\hat\theta_1 +\partial_t\hat\theta_1\right]\ ,
\end{equation}
where $v_0=\frac{\hbar\partial_x\theta_0}{m}$ is the condensate velocity (we consider a quasi-1D condensate for simplicity), while the equation for $\hat\theta_1$
decouples and is mathematically equivalent to the Klein-Gordon equation for a massless, minimally coupled scalar field
\begin{equation} \label{kgam} \frac{1}{\sqrt{-g}}\partial_\mu (\sqrt{-g}g^{\mu\nu}\partial_\nu  \hat \theta_1) =0 \  \end{equation}
in the background metric
\begin{equation} \label{acm} ds^2=\frac{n}{mc_s}\left( -(c_s^2-v_0^2)dt^2+2v_0dtdx +dx^2+dy^2+dz^2\right)\ .
\end{equation}
Eq. (\ref{kgam}) is the equation Hawking used to quantize a scalar field in the Schwarzschild background. 
The acoustic metric (\ref{acm}) possesses an acoustic horizon located at $|v_0|=c_s$ with surface gravity $\kappa=\frac{1}{2c_s} \frac{d(c_s^2-v_0^2)}{dx}|_{hor}$, and unlike its gravitational analogue it does not need to have the interior (supersonic) region to terminate into a (nasty) spacelike singularity.
Since Hawking's analysis only relies on the kinematical properties of the horizon, event or acoustic, the prediction that in the formation of an acoustic black hole a thermal flux of phonons at the temperature (\ref{hawktempan}) will be produced is then straightforward. 

The experimental realization of an acoustic black hole in a BEC was carried out for the first time in in \cite{lahav}. Because of their extremely low temperatures (of the order of $T_C\sim 100 nK$, much below $T_{CMB}$) they  offer much more favourable experimental conditions for the detection of the Hawking effect with respect to gravity. Indeed, acoustic black hole configurations can be constructed for which $T_H^{an}\sim 10 nK$, only one order of magnitude below $T_C$. Nevertheless, the signal is tiny and competing effects (the bigger background temperature, quantum noise, etc..) still easily overcome the Hawking flux of the created phonons making it practically unobservable. 

\section{One step forward: the Hawking effect in density correlations in BECs} 

Another experimental advantage of having to do with acoustic black holes instead of gravitational ones
is that acoustic horizons, unlike event horizons, allow measurements to be performed on both sides, the exterior (subsonic) and the interior (supersonic) regions. This is very important because we can then try to characterize experimentally the Hawking effect through the pair production process responsible for it. 
This was done in \cite{pra}, to which we refer for more details.   

In \cite{pra} the phase fluctuation $\hat \theta_1$ was treated as a 2D conformal scalar field propagating in the $(t,x)$ section of the acoustic metric (\ref{acm}), for which the two-point function $\langle \hat \theta_1 \hat \theta_1 \rangle $ is known analytically \footnote{This approximation only slightly overestimates the flux of the emitted phonons.}.  This powerful result allowed us to construct in closed form, using (\ref{df}), the equal time normalized density-density correlation function in the Unruh state  (the state appropriate to describe black hole evaporation). Focussing in particular on points situated on either side with respect to the acoustic horizon, we got the characteristic stationary behaviour 
\begin{equation}\label{corrsign}
 G^{(2)}_{BH}\equiv \frac{1}{n_0^2}\langle \hat n_1\hat n_1\rangle \sim \kappa^2 \cosh^{-2}[\frac{\kappa}{2}(\frac{x}{v+c_l}-\frac{x'}{v+c_r})]\ .
\end{equation}
Here $v<0$, i.e. the flow is from right to left while $x=0$ is the acoustic horizon dividing the right ($x<0$) subsonic and left ($x>0$) supersonic regions.  The formation of the horizon triggers the spontaneous creation of pairs of phonons (of equal and opposite frequencies, as required by energy conservation) on both sides of the horizon. In the lab they propagate along the lines  $x=(v+c_l)t\ (<0)$ (partner), $x'=(v+c_r)t$ (Hawking quanta), see Fig. 2. \footnote{Although we are considering its hydrodynamic/relativisistic approximation, the system we started with is nonrelavistic and velocities add up following Galileo's law.} This happens for all $t$  and leads to a stationary signal correlating points $x,x'$ such that 
\begin{equation}\label{hawkpeak}
\frac{x}{v+c_l}=\frac{x'}{v+c_r}\ .
\end{equation}
 The correlator (\ref{corrsign}) has a peak exactly along this line: 
this signal is a clear manifestation of the pair-production process at the base of the Hawking effect. 
Quantitative estimates based on existing experiments suggest the the magnitude of this peak, of the order of 
$10^{-3}$, is small but not negligible.

\begin{figure}
\centering \includegraphics[angle=0, height=1.7in] {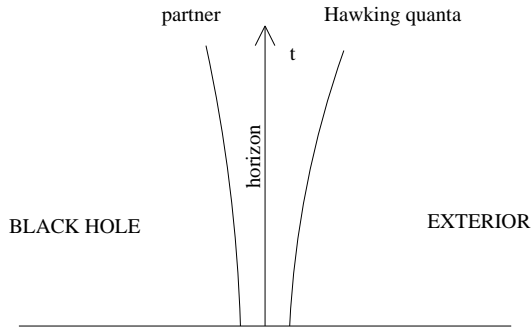}
\caption{Pair-creation (Hawking quanta - partner) from the black hole horizon. }
\end{figure}

One can object that this result and the thermal flux prediction, being 
based on the gravitational analogy, are subject to a  `transplanckian' problem analogous to that in  
gravity. Indeed, sound waves propagating from the near horizon region get exponentially redshifted, and this implies that the physics of the Hawking effect  is in principle sensitive to microscopic scales, in particular those smaller than $\xi$. In BECs we can study the `transplanckian' problem by going back 
to the  basic equations (\ref{gp}) and (\ref{bdg}), valid at all scales (we do not have the analogous equations in gravity). 

The result (\ref{corrsign}) served both as a motivation and a guide for a subsequent numerical analysis performed within the full theory using Montecarlo techniques \cite{njp}. The results, shown in Fig. 3,
confirmed the existence of the Hawking peak in the microscopic theory.  A quantitative
comparison of the height and width of the peak with (\ref{corrsign}) showed excellent agreement with the hydrodynamical prediction in the regime $\kappa\ll \frac{1}{\xi}$. Outside this regime (which in the gravity context would require knowledge of a quantum theory of gravity) acoustic black holes still emit at an approximately thermal rate (up to a certain cutoff frequency $w_{max}\sim \frac{1}{\xi}$ \cite{mapa}) and the location of the Hawking peak is still given by (\ref{hawkpeak}) (for analytical results in this regime, see \cite{bcfmr} and references therein).

\begin{figure}
\centering \includegraphics[angle=0, height=2.5in] {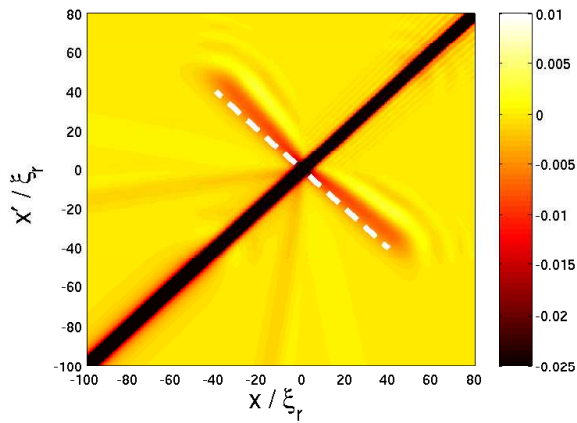}
\caption{Numerical results for the density correlator confirming the existence of the Hawking peak (red tongue perpendicular to the main diagonal) in the microscopic BEC theory. }
\end{figure}

A nice way to explain why there is no transplanckian problem in BECs is depicted in Fig. 4. By tracing backwards in time the trajectories of the Hawking quanta and partners,
once the near horizon region is reached microscopic physics effects change the phonon's dispersion relation which becomes `superluminal'. This causes the two trajectories to stop hovering close to the horizon (where, instead, they would remain forever in the hydrodynamical/relativistic approximation) and enter the interior region. The resulting blueshift along the trajectories is now finite, of the order of $\frac{1}{\xi}$ (we recover the infinite blueshift in the hydrodynamcal $\xi \to 0$ limit). 
\begin{figure}
\centering \includegraphics[angle=0, height=2.0in] {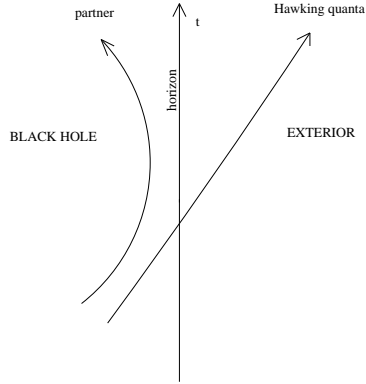}
\caption{Because of short-distance effects  (scales smaller than $\xi$)  Hawking quanta and partners
appear to be produced in the interior of the acoustic black hole, thus resolving the transplanckian problem
in BECs.}
\end{figure}

Another important result of the numerical analysis in \cite{njp} is that the Hawking peak 
of Fig. 3 is still visible even when a nonzero thermal background (with a temperature bigger than $T_H$) is present. Thus, unlike the emitted thermal flux (which is easily masked by the background ) the correlator signal (\ref{corrsign})  is the looked for 'smoking gun' of the Hawking effect in the experimental search.

\section{Acoustic white holes in BECs: another experimental signature of the Hawking effect}

White holes, time reversal of black holes, have not captured much interest in gravity. They expel to the exterior everything that lies inside their horizon, exactly the opposite behaviour of black holes. They `emerge' from an initial singularity, a kind of big-bang singularity, which prevents
physical predictions about them to be made until we'll find a way to deal with singularities in the quantum theory. 

This problem is not present in the fluid case, and acoustic black and white holes can be studied on the same footing. A schematic representation of an acoustic white hole is given in Fig. 5, and is simply obtained  by reversing the sign of the fluid velocity $v_0$ in Fig. 1 (and also in the acoustic metric (\ref{acm})).  An acoustic white  hole is represented by a supersonic fluid decelerating until it becomes subsonic, whereas the acoustic black hole is the other way around, namely an accelerating fluid turning from subonic to supersonic. In both cases, the acoustic horizon lies where the fluid velocity equals the sound speed.

 \begin{figure}
\centering \includegraphics[angle=0, height=1.5in] {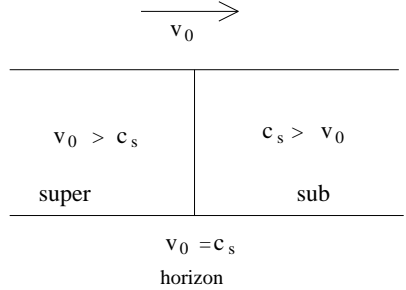}
\caption{Schematic representation of an acoustic white hole in a fluid, obtained from Fig. 1 by reversing
the sign of the fluid velocity $v_0$. }
\end{figure}

Spontaneous pair production in stationary backgrounds requires the existence of negative energy states. These exist in supersonic regions of both acoustic black holes and white holes, indicating that an analog Hawking effect exists in both configurations and not just in black holes \footnote{Perhaps surprisingly, such an effect exists also for an everywhere supersonic fluid with no acoustic horizon, see \cite{bcfmr}. What seems to depend on the existence of an acoustic horizon is the thermal character of the created flux.}.  Its features in white holes are however totally different. Because of the time-reversal property with respect to the black hole case (Fig. 2), the trajectories in Fig. 6 show that Hawking quanta and partners now accumulate along the horizon, and they do this  with a diverging blueshifted frequency.  This implies that hydrodynamical white hole horizons (as well as relativistic ones) are unstable: the Hawking effect in white holes crucially depends on the details of the short-distance physics.

 \begin{figure}
\centering \includegraphics[angle=0, height=1.5in] {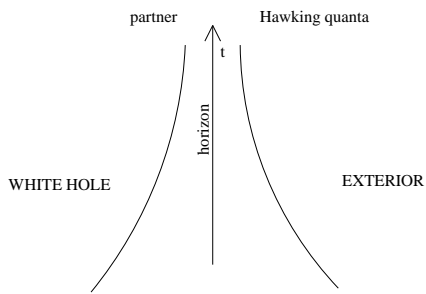}
\caption{Trajectories of Hawking quanta and partners in acoustic white holes. }
\end{figure}

The `superluminal' modification of phonon's dispersion relation in BECs causes (as in the black
hole case)  departure from the hydrodynamical picture in Fig. 6 when a maximum frequency $\sim \frac{1}{\xi}$ is reached, allowing both trajectories to enter the horizon (Fig. 7).

These arguments indicate that the main Hawking signal in BEC acoustic white holes takes 
place inside the horizon. 
Analytical results derived within the linear Bogoliubov theory show that for $x, x'<0$, i.e. inside the horizon,
the density-density correlator has the typical oscillating behaviour \cite{wh}
\begin{equation}\label{corrsignwh}
G^{(2)}_{WH}(x,x')\sim \left[A\cos p_0(x+x') + B\sin p_0(x+x') + C\cos p_0(x-x')\right]I_{\epsilon}\ ,
\end{equation}
with $p_0$ the nontrivial zero mode ($\sim \frac{1}{\xi}$) of the `superluminal' Bogoliubov dispersion (present only in the supersonic region), and the overall amplitude 
$I_{\epsilon}=\int_{\epsilon}\frac{dw}{w}$, formally infrared divergent,  is regulated by introducing the low-frequency cutoff $\epsilon=\frac{1}{t}$ where $t$ measures time elapsed from horizon formation. 
A closer inspection shows \cite{cpf} that the two-point function in the relevant low-frequency regime 
factorizes as the product of the same function $\phi_0$ evaluated at the points $x$ and $x'$, typical of an
emerging classical behaviour. It describes the emission of a zero-frequency undulating wave away from the horizon with a macroscopically growing amplitude  (i.e. the $\frac{1}{w}$ term in $I_{\epsilon}$)
fixed by the low-frequency spectrum of the spontaneously produced Hawking phonons. It is worth to mention that undulations were observed in water tanks experiments in \cite{silke}, but their connection with the Hawking effect was not pointed out.

\begin{figure}
\centering \includegraphics[angle=0, height=2.0in] {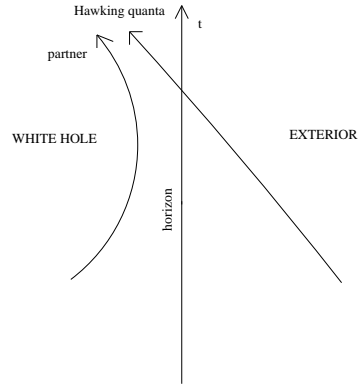}
\caption{Short-distance physics  (scales smaller than $\xi$) cuase Hawking quanta and partner
to enter the white hole horizon instead of accumulating along the horizon on both sides (Fig. 6).}
\end{figure}

\section{Conclusions and future prospects} 

We have described how the analog Hawking radiation can be measured in Bose-Einstein condensates by a direct measurement of the density correlations between the spontaneously produced pairs of phonons (Hawking quanta and partners) on both sides of the acoustic horizon (such a measurement would be impossible to perform in gravity). 
This provides the looked for 'smoking gun' of the Hawking effect and makes these type of measurements very promising for 
an experimental verification of the Hawking effect in the near future.
Methods to amplify this signal have been proposed \cite{cornell} by letting the condensate to expand freely 
after the acoustic black hole is formed.

Finally, it is worth to mention that density correlation measurements between atoms velocities were performed recently in \cite{chris} to observe the creation of correlated excitations with equal and opposite momenta in homogeneous condensates undergoing a sudden variation of their trapping potential. While this effect can be interpreted  as the analog  of cosmological particle creation,  this technique will prove useful to test the Hawking effect in stationary supersonic flows.

\acknowledgments
This work is part of a research developed within the Project ''Acoustic black holes'' of Centro Fermi, in collaboration with Roberto Balbinot.

\end{document}